\documentclass[twocolumn,showpacs,preprintnumbers,amssymb,aps]{revtex4}
\usepackage{dcolumn}
\usepackage{bm}
\usepackage{graphicx}
\usepackage{amsmath}

\usepackage{changes}

\begin{document}

\title{Universal and robust dynamic decoupling controls for zero-field magnetometry by using molecular clock sensors}

\author{Jiawen Jiang and Q. Chen\footnote{E-mail:qchen@hunnu.edu.cn}}
\affiliation{ Key Laboratory of Low-Dimension Quantum Structures and Quantum Control of Ministry of Education, \\
Synergetic Innovation Center for Quantum Effects and Applications, Xiangjiang-Laboratory and Department of Physics, \\
Hunan Normal University, Changsha 410081, China
}

\begin{abstract}
Color centers in diamond and silicon carbide (SiC), and molecular spins through a host matrix control are promising for nanoscale quantum sensing because they can be optically addressable, coherently controllable, and placed proximate to the targets. However, large transverse zero-field splitting (ZFS) is often inevitable due to their intrinsic symmetry and/or the high local strains of the host matrix. Although spin coherence can be extended due to magnetic noise-insensitive clock transitions at a vanishing magnetic field, the eigenstates of these sensors are not sensitive to weak magnetic signals in the linear order. We address this challenge by employing a combination of radio-frequency (RF) field driving along the NV orientation and microwave (MW) dynamic decoupling pulse sequences. RF driving can effectively mitigate the transverse ZFS effect and enhance the NV center’s sensitivity to AC magnetic field signals. This combination not only suppresses environmental noise but also enables quantum frequency mixing between the transverse ZFS and the signal. It also offers the potential to detect weak AC signals at intermediate and high frequencies with high resolution, a task difficult to achieve using conventional methods.
\end{abstract}

\maketitle
\section{Introduction}
Color centers embedded in diamond and silicon carbide (SiC) \cite{Norman2021} have shown great success in quantum information processing. For example, nitrogen vacancy (NV) centers \cite{doherty2013nitrogen,degen2017quantum,sensitivity,wu2016diamond} in diamond can be optically initialized, read out, and coherently controlled, providing a promising quantum sensing platform due to their nanoscale resolution and biocompatibility \cite{lovchinsky2016nuclear,schmitt2017submillihertz,aslam2017nanoscale,shi2015single}. Different kinds of defects in SiC \cite{luo2023fabrication,Li2022temperature} are also used for high-sensitivity quantum metrology of magnetic fields, electric fields, local strain fields, and temperature \cite{niethammer2016vector,wolfowicz2018electrometry,zhou2017self,falk2014electrically,wang2023magnetic}. Recently, increasing attention has been paid to chemically synthesized systems such as molecular color center \cite{bayliss2020optically,laorenza2021tunable,bayliss2022enhancing} because they are zero-dimensional with Angstrom-scale precision, high qubit tunability, and flexibility. One of the major challenges limiting the quantum sensing applications of these color centers is the shortened coherence time caused by the noisy environment.

A large transverse zero-field splitting (ZFS) is often inevitably introduced due to symmetry and local strain.
Recent studies have shown that this transverse ZFS enables magnetic noise-insensitive clock transitions near zero external magnetic field, thereby increasing the coherence time of NV centers in diamond \cite{nanodiamonds,Dolde2011Electric,2016Competition}, defects in SiC \cite{miao2020universal,onizhuk2021probing,zhu2021theoretical}, and molecular spins \cite{bayliss2022enhancing,shiddiq2016enhancing}. Many efforts are devoted to color center magnetometry in the zero- and low-field regimes \cite{Cerrillo2021Three,vetter2022zero,wang2022zero,saijo2018ac,zheng2019zero,lenz2021magnetic} because a bias magnetic field can introduce undesirable effects when analyzing the intrinsic magnetic properties of materials and molecular structures. However, zero-field magnetometry using these sensors (hereafter referred to as molecular clock sensors) is challenging due to their insensitivity to weak magnetic signals. On the other hand, the second-order magnetic field noise from the nuclear spin bath could be significant for electrically neutral molecular spins \cite{bayliss2022enhancing}, and color centers in solid-state hosts are susceptible to surface electric noise \cite{romach2015spectroscopy,Chrostoski2018electric,myers2017double}. Electric noise could be the dominant noise source, considered responsible for the short decoherence time of shallow NV centers \cite{2016Competition,Kim2015Decoherence} or in nanodiamonds \cite{gardill2020fast}. Previous efforts to improve the coherence time of these color centers have mainly focused on magnetic noise.

%%%%%%%%%%%%%%%%%%%%%%%%%%%%%
\begin{figure*}[ht]
\center\includegraphics[scale=0.55]{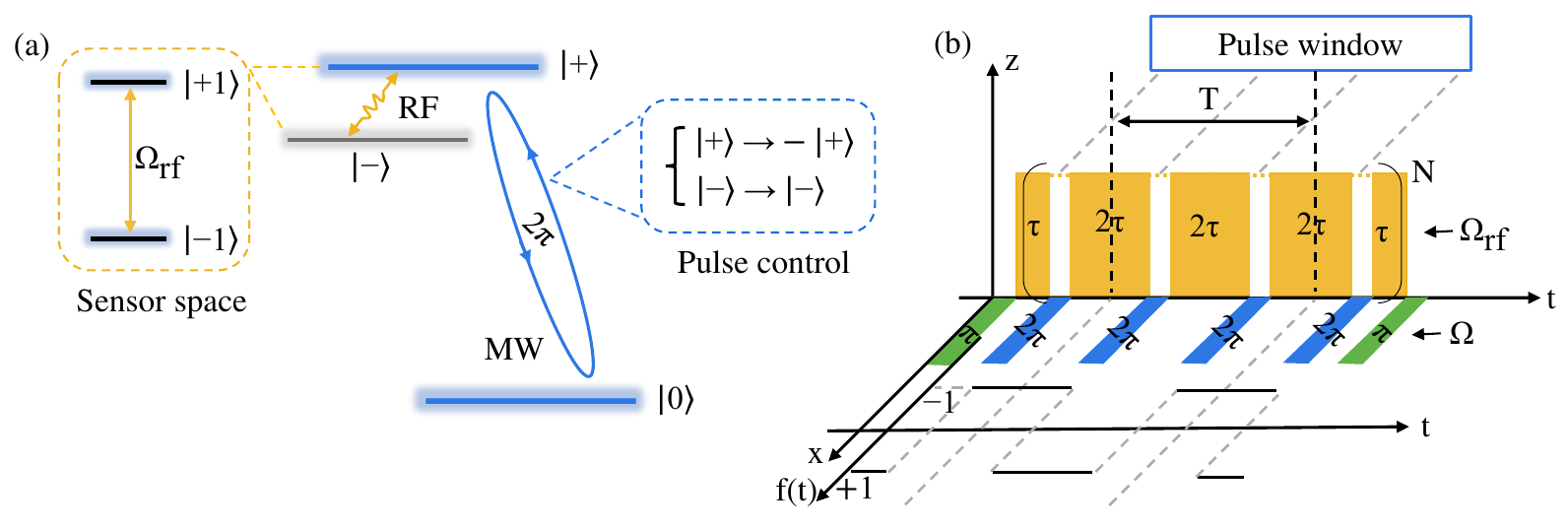}
\caption{(a) The eigenstates of the molecular clock sensor and $|\pm\rangle$ are splitted by transverse strain or electric field $\sim2E_{x}$ under zero-field. Clock sensor is working in state space of $|\pm1\rangle$. The continuous RF field is applied along the sensor axis and the MW field is applied along the transverse direction. A $\pi$ microwave pulse initializes the clock sensor from state $|0\rangle$ to $|+\rangle$ state, when $2\pi$ pulse has $|+\rangle\rightarrow-|+\rangle$, which induces $|\pm1\rangle\rightarrow-|\pm1\rangle$. (b) The basic control schematic of our scheme, in which the time $\tau$ is related to the period of filter function $f(t)$. }
\label{pulse}
\end{figure*}
%%%%%%%%%%%%%%%%%%%%%%%%%%%%%
We address these challenges by using a combination of a continuous RF driving along the NV orientation and MW pulse sequences to have universal and robust dynamic decoupling (DD) controls for zero field magnetometry of molecular clock sensors. Instead of a static bias magnetic field, continuous RF driving is applied along spin's orientation to resonate with the energy difference induced by the transverse ZFS, which makes the sensor sensitive to weak magnetic signals. Moreover, the combination of continuous RF driving and DD pulses inherits the robustness to nearly all the decoherence channels in the sensor, such as electric noise, magnetic fluctuations and control errors of RF driving and DD pulses. Therefore, its coherence time could be increased more than 3 orders of magnitude for different kinds of molecular clock sensors, such as color centers in diamond, SiC and optical addressable molecular spins. Our method can efficiently decouple the sensor from environmental noise and create narrow-band filters to sense nearby weak AC magnetic fields. Our technique enables detection over a wide AC frequency range (kHz to GHz) because the transverse ZFS generates an intermediate frequency scale, which is further controlled by suitably modulated microwave control fields.

\section{Control methods}
We consider a molecular clock sensor with a spin triplet ground state, where a large transverse ZFS is induced by altering the system symmetry of the matrix, local strains, or applying an electric field \cite{iwasaki2017direct} when the static magnetic field is weak enough to be negligible. The level splitting of the ground spin-triplet state can be described by the following effective Hamiltonian
\begin{eqnarray}\label{total}
H_{s}&\approx&DS_{z}^{2}+(E_{x}+\delta _{E})(S_{x}^{2}-S_{y}^{2}),\notag
\end{eqnarray}
in which $D$ is the uniaxial ZFS, $E_{x}$ is the transverse ZFS which can be determined precisely by using Ramsey interferometry of the $|+\rangle\leftrightarrow|-\rangle$ transition. Eigenstates are given by $|0\rangle$ and $|\pm\rangle$ states ( $|\pm\rangle=\frac{1}{\sqrt{2}}(|+1\rangle\pm|-1\rangle)$) as shown in Fig. 1(a). There are so-called clock transitions between $|0\rangle$ and $|\pm\rangle$ states, in which transition frequencies are given by $D\pm E_x$, insensitive to first-order magnetic noise, namely $df/dB\rightarrow0$. Therefore, the spin coherence is protected from first-order magnetic noise, and we can assume that system decoherence is dominated by electric noise, strain noise, and/or second-order magnetic fluctuations $\delta _{E}$, detailed explanations is in Appendix B.

A sufficiently strong RF field $H_r=\Omega_{rf}(1+\eta_{r})\cos(\omega_{rf}t)S_z$ is continually driven along the sensor's orientation $z$-direction to be resonant with the energy between eigenstates $|\pm\rangle$ with $\omega_{rf}=2E_{x}$ and Rabi frequency $\Omega_{rf}$, and $\eta_{r}$ is the relative control error. Dynamics of evolution could be characterized as a corresponding two-level system working in $|\pm1\rangle$ subspace, thus we can rewrite the Hamiltonian of the system under the RF driving with $H_0=DS_{z}^{2}+E_{x}(S_{x}^{2}-S_{y}^{2})$ as
\begin{eqnarray}\label{free}
H_{rf}\approx\Omega_{rf}(1+\eta_{r})\sigma_{z}+2\delta_E\sigma_{x},
\end{eqnarray}
where the Pauli operators $\sigma_{z}=\frac{1}{2}(|+1\rangle\langle+1|-|-1\rangle\langle-1|)$ and $\sigma_{x}=\frac{1}{2}(|+1\rangle\langle-1|+|-1\rangle\langle+1|)$. The strong Rabi frequency $\Omega_{rf}$ ($\delta_{E}\ll\Omega_{rf}$) in effect suppresses the electric noise and energy gap is formed between spin states $|\pm1\rangle$ which are the basis of our sensor. But it brings in control errors ($\eta_{r}$) in real experimental situations.

Dynamical decoupling is a well-established technique that employs $\pi$ pulses to invert the spin states of the sensor, thereby effectively filtering out unwanted environmental noise. In our quantum system, we use a $2\pi$ MW pulse to have $|+\rangle\rightarrow-|+\rangle$ state while state $|-\rangle$ is unaffected (approximated as a $\pi$ pulse transition in $\{|+1\rangle,|-1\rangle\}$ subspace). Microwave pulses are applied for the state transition between $|0\rangle$ and $|+\rangle$ with the driving frequency $\omega=D+E_{x}$, the system's rotating frame Hamiltonian becomes
\begin{eqnarray}\label{d1}
H_{d}&\approx&\frac{\Omega(1+\eta_{m})}{2}(|0\rangle\langle +|+|+\rangle\langle 0|)| \\
& &+\frac{\Omega_{rf}(1+\eta_{r})}{2}(|-\rangle\langle +|+|+\rangle\langle -|)\notag\\
& &+\delta_E(|+\rangle\langle +|-|-\rangle\langle -|).\notag
\end{eqnarray}
When the Rabi frequency of MW pulses are far larger than the RF driving $\Omega\gg\Omega_{rf}\gg\delta_E$, the length of a pulse is approximately defined as $T_{p}\approx\frac{2\pi}{\Omega}$ ($\Omega T_{p}=2\pi$). $\eta_{m}$ is the relative control error. Applying such a $2\pi$ pulse, the state $|+\rangle$ evolves as $\exp\left[-i\int_{0}^{T_{p}}H_{d}dt\right]|+\rangle=\exp(-i\pi\sigma_{x}^{\{|0\rangle,|+\rangle\}})=-|+\rangle$, where $\sigma_{x}^{\{|0\rangle,|+\rangle\}}$ is the Pauli operator in $\{|0\rangle,|+\rangle\}$ subspace. So $\pi$ pulse within $\{|+1\rangle,|-1\rangle\}$ subspace could be realized. Thus, there is a trade-off that a stronger RF Rabi frequency suppresses electric noise better but brings in more control errors in sequences because of state leakages to the $|-\rangle$ state. Different kinds of DD sequences modulated by phases \cite{vetter2022zero,Genov2014correction,genov2017arbitrarily,Ryan2010Robust}, such as LDD illustrated in Appendix D, are proposed to show improved robustness to the driving field noise.

Now one can obtain intuition about our DD control mechanism by considering a molecular sensor working in the $\{|+1\rangle,|-1\rangle\}$ basis. At zero field, a large transverse ZFS can eliminate the effect of first-order magnetic noise, and the sensor is limited by fluctuations $\delta_{E}$, which are determined by electric and/or second-order magnetic noise. Strong continual RF driving (assuming $\delta_{E}\ll\Omega_{rf}$) can overcome the effect of such fluctuations, form the energy gap between states $|+1\rangle$ and $|-1\rangle$, and introduce driving field fluctuations. The system's noise is dominated by its second-order term, approximately $\sim\delta_{E}^2/\Omega_{rf}$. By utilizing phase-fixed geometric controls in the ${|0\rangle,|+\rangle}$ subspace, $\pi$ pulses within the ${|+1\rangle,|-1\rangle}$ subspace can cancel second-order terms, and different kinds of DD sequences are used to improve performance. Therefore, the coherence time is mainly determined by the third-order term $\sim2\delta_{E}^3/\Omega_{rf}^2$ when control errors are negligible. The specific mechanism and pulse sequences are presented in Fig. \ref{pulse}(a) and \ref{pulse}(b). Therefore, our mechanism employs three methods, including clock transition, continuous driving field, and sequence of pulses, to protect spin coherence from environmental noise and control errors. Significant prolongation of the spin coherence in the ${|+1\rangle,|-1\rangle}$ basis could be possible, indicating sufficient manipulation fidelity and coherence resources for quantum sensing purposes. We use zero-field AC magnetometry as an example to demonstrate the efficiency of our method.

\section{Sensing applications}
We assume an AC signal along the direction of the sensor's axis defined as $z$ direction and $H_s=g_{ac}\cos(\omega_{ac}t)S_z$, where $g_{ac}$ and $\omega_{ac}$ are the weak coupling strength and frequency of AC fields, respectively. Go to rotating frame with respect of $H_{0}=(D+E_z)S_{z}^{2}+E_{x}(S_{x}^{2}-S_{y}^{2})$, if we have $|\omega_{ac}-2E_{x}|\ll E_{x}\ll |\omega_{ac}+2E_{x}|$ and $\frac{1}{2}\Omega_{rf} \ll 2E_{x}$, after the rotating-wave approximation the effective Hamiltonian of the whole system is given by
\begin{widetext}
\begin{eqnarray}\label{eff}
H_{eff}&\approx& \left(
            \begin{array}{ccc}
              \frac{1}{2}(1+\eta_{r})\Omega_{rf}+  \frac{1}{2}g_{ac}\cos(\Delta\omega t) &  \frac{\sqrt{2}}{4}(1+\eta_{m})\Omega(t) & \delta_E \notag \\
               \frac{\sqrt{2}}{4}(1+\eta_{m})\Omega(t) & 0 &  \frac{\sqrt{2}}{4}(1+\eta_{m})\Omega(t) \\
              \delta_E &  \frac{\sqrt{2}}{4}(1+\eta_{m})\Omega(t) & -\frac{1}{2}(1+\eta_{r})\Omega_{rf}-\frac{1}{2}g_{ac}\cos(\Delta\omega t) \\
            \end{array}
          \right),
\end{eqnarray}
\end{widetext}
where $\Delta\omega=\omega_{ac}-2E_{x}$. Our method works when $|\omega_{ac}-2E_{x}|\ll|\omega_{ac}+2E_{x}|$ and we assume $\omega_{ac},E_{x}>0$, the detailed calculations are given in Appendix B.

When the state is initialized to $|+\rangle$ state, the AC signal is accumulated during the RF continual resonant driving. Dynamics of free evolution could be characterized as a corresponding two-level system working in $|\pm1\rangle$ subspace, and we can rewrite the Hamiltonian of free evolution as
\begin{eqnarray}\label{free}
H_{free}\approx\Omega_{rf}(1+\eta_{r})\sigma_{z}+g_{ac}\cos(\Delta\omega t)\sigma_{z}+2\delta_E\sigma_{x}.
\end{eqnarray}
DD pulses are used to invert spin states in $|\pm1\rangle$ subspace through the control of transition between states $|0\rangle$ and $|+\rangle$ and the effective Hamiltonian of whole system with RF and MW DD controls is as follows
\begin{eqnarray}\label{Ht}
H(t)&\approx&f(t)[\Omega_{rf}(1+\eta_{r})\sigma_{z}\\
&+&g_{ac}\cos(\Delta\omega t)\sigma_{z}+2\delta_{E}\sigma_{x}]\notag,
\end{eqnarray}
where $f(t)$ is the response filter function flipping between $+1$ and $-1$ corresponding to the inversion of MW pulses, see Fig. \ref{pulse}(b). One can have periodic filter function to be on resonance with $\Delta\omega$, so that phases are accumulated between states $|+1\rangle$ and $|-1\rangle$ in free evolution times corresponding to target signals. Thus, if the target frequency is resonant with modulation frequency, i.e.,
\begin{eqnarray}\label{Ht}
\Delta\omega=\omega_{ac}-2E_{x}=\pi/(2\tau),
\end{eqnarray}
we have an accumulated phase $\eta(t)\equiv\int_{0}^{t}g_{ac}|\cos(\Delta\omega t)|$. The AC signal then could be read out by calculating the population of the NV sensor on the $|+\rangle$ state.

%%%%%%%%%%%%%%%%%%%%%%%%%%%%%
\begin{figure}[htp]
\center\includegraphics[scale=0.58]{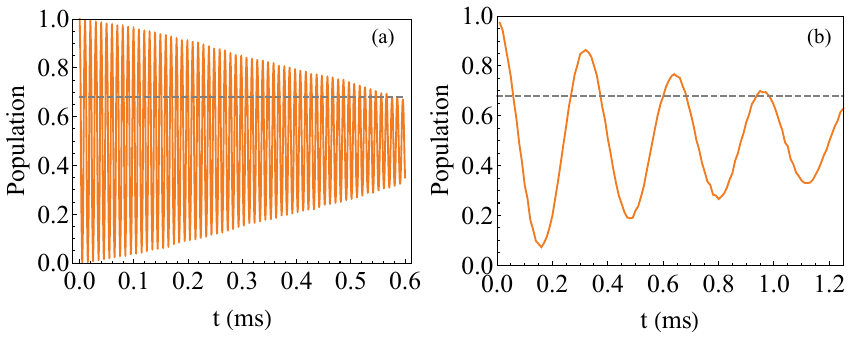}
\caption{The signal is given by the population evolution of the NV center in the $|+\rangle$ state. (a) There is only continuous RF field control when electric noises, RF control errors and magnetic field fluctuations are included. (b) The DD MW pulses are applied to prolong coherence time, where the Rabi frequency of MW pulse is $\Omega=(2\pi)$ $40$ MHz, the coupling strength of weak AC field $g_{ac}=(2\pi)$ $5.0$ kHz, and the resonance condition $\Delta\omega=(2\pi)$ $0.1$ MHz.}
\label{bulk}
\end{figure}
%%%%%%%%%%%%%%%%%%%%%%%%%%%%%

\section{Numerical simulations}
We show the performances of RF and MW DD controls by averaging the results results over 1200 simulations using exact numerical methods. The system operates in a three-state basis affected by electric noise $\delta_E$ (constrained by $T_2^*$) \cite{trusheim2014scalable} and uncorrelated power fluctuations of the driving fields $\eta_{m}$, $\eta_{r}$ following an Ornstein-Uhlenbeck process \cite{wang1945theory,gillespie1996mathematics,gillespie1996exact}. The $\eta_{m}$ and $\eta_{r}$ are set as $0.5\%$ in all of our simulations \cite{miao2020universal,genov2020adiabatic,Casanova2020Continuous,Cohen2017,Aharon2019}. The effective Hamiltonian of the whole system for the simulations is given by Eq. (3). Electric fluctuation $\delta_{E}$ is determined by dephasing time $T_{2}^{*}$ which could be in the order of $\mu$s to tens of  $\mu$s in current typical experimental reports of NV centers, defects in SiC and optical addressable molecular spins in different environments.

\begin{table}[h]
\setlength{\tabcolsep}{5.5pt}
\caption{The extended coherence time $T_{2}$ varying with expectation dephasing time $T_{2}^{*}$ under different RF driving control and pulse sequences. The Rabi frequency of MW pulse is $\Omega=(2\pi)$ $40$ MHz and the resonance condition $\Delta\omega=(2\pi)$ $0.1$ MHz.
}
\begin{center}
\begin{tabular}{m{1.7cm}<{\centering}m{1.2cm}<{\centering}cccc }
\toprule
$\Omega_{rf}$ & \textrm{Pulse} &  & $T_{2}^{*}$ & ($\mu$s)  \\
$[(2\pi)$ MHz] &
 \textrm{control}& 0.9 & 1.8 & 5.4 & 10\\
\hline
 & \textrm{None} & 15.1 & 40.3 & 83.3 & 169.3\\
2.0& DD & 26.9 & 138 & 2628 & 4297\\
 & LDD & 28.6 & 138 & 5150 & 11540\\
\hline
 & \textrm{None} & 24.9 & 40.1 & 47.1 & 44.8\\
4.0& DD & 48 & 280 & 380 & 525\\
 & LDD & 48 & 585 & 5012 & 7015\\
\hline
 & \textrm{None} & 27.3 & 34.5 & 33.5 & 36.6\\
6.0 & DD & 61 & 106 & 178 & 172\\
 & LDD & 98 & 1050 & 2553 & 3229\\
\botrule
\end{tabular}
\end{center}
\end{table}
%%%%%

\begin{figure}[htp]
\center\includegraphics[scale=0.7]{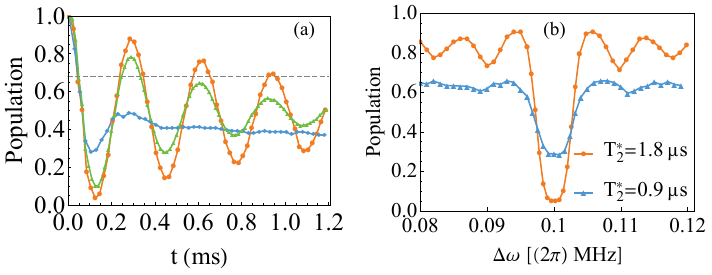}
\caption{The signal is given by the population evolution of the NV center in the $|+\rangle$ state in which electric noises, control errors and magnetic field fluctuations are included. The Rabi frequency of MW pulse is $\Omega=(2\pi)$ $40$ MHz, the coupling strength of weak AC field $g_{ac}=(2\pi)$ $5.0$ kHz, and the resonance condition $\Delta\omega=(2\pi)$ $0.1$ MHz. (a) The expectation coherence time $T_{2}^{*}=1.8$ $\mu$s (orange and green line) and $T_{2}^{*}=0.9$ $\mu$s (blue line) are considered. Compare to $\Omega_{rf}=(2\pi)$ $4$ MHz (green line), a larger RF frequency of $\Omega_{rf}=(2\pi)$ $6$ MHz (orange line) shows better robustness to noises. (b) The spectrum signals are given with $\Omega_{rf}=(2\pi)$ $6$ MHz and the other parameters are the same as (a).}
\label{spectrum}
\end{figure}

Take an NV center with $E_x=(2\pi) 300$ kHz as an example, suppose that magnetic noise are the dominated noise, clock transitions protect spin coherence leading to $T_{2}^{*}=35$ $\mu$s \cite{2016Competition} at zero field. A single RF driving field $\Omega_{rf}=(2\pi)$ $100$ kHz increases the coherence time $T_2$ about an order due to $\delta_{E}\ll\Omega_{rf}$ and MW DD controls induce a substantial increase of $T_2$ to about 1 ms, see Fig. \ref{bulk}(b), here the control errors are ignored. But if we consider a molecular clock sensor working in noisy environment when fluctuations $\delta_{E}$ are dominated, e.g., shallow NV centers or molecular color centers, a large Rabi frequency driving of RF field is necessary to have $\delta_{E}\ll\Omega_{rf}$ and control errors are not negligible. Fortunately, our technique inherits the robustness against control errors because of the included well-developed DD sequences (LDD sequences are used as an example in our simulations). As we discussed, there is a trade-off that a stronger RF Rabi frequency suppresses electric noise better but brings in more control errors, in order to estimate the optimal Rabi frequency driving of our method, we show several cases when $T_2^*<10$ $\mu$s, detailed simulation results are presented in Table 1 and Appendix D. For instance, we evaluate the decoherence time extended to $T_{2}=585$ $\mu$s with RF Rabi frequencies $\Omega_{rf}=(2\pi)$ $4$ MHz when $T_{2}^{*}=1.8$ $\mu$s \cite{trusheim2014scalable}, and a larger RF driving with $\Omega_{rf}=(2\pi)$ $6$ MHz induces a longer decoherence time $T_{2}=1050$ $\mu$s shown in Fig. \ref{spectrum}(a), but our simulation shows there's no more benefit if we continue to increase RF driving. Notice that we assume that the population relaxation time of the sensors are long enough for simplicity.
%%%%%%%%%%%%%%%%%%%%%%%%%%%%%
%%%%%%%%%%%%%%%%%%%%%%%%%%%%%

The molecular sensor with a large transverse ZFS not only protect the spin coherence from environmental noises and control errors but also plays a role as a frequency mixer to the target AC field signals, and we show the high resolution spectrum of resonant frequency ($\Delta\omega=\omega_{ac}-2E_{x}$) in Fig. \ref{spectrum}(b). The recent reported transverse ZFS can be several MHz for shallow NV centers in diamond or more than 10 MHz in nanodiamonds \cite{nanodiamonds} due to local strains near the surface, could reach tens of MHz or be larger than 100 MHz for divacancy or NV centers in SiC \cite{luo2023fabrication}, and GHz for molecular spins \cite{bayliss2022enhancing}. The transverse ZFS could be adjusted by altering the system symmetry of the matrix, local strains or applying an electric field \cite{iwasaki2017direct}, which makes detectable frequencies range from hundreds of kHz to GHz. This allows for a broadened sensing bandwidth and the analysis of electric field noise even with a short dephasing time. Our scheme also provides probability for some frequencies which are difficult to reach by using conventional methods and bring in new sensing applications.

\section{Conclusion and discussion}
Compared to current schemes for zero-field magnetometry \cite{wang2022zero,saijo2018ac,zheng2019zero,lenz2021magnetic}, which use a weak bias magnetic field or strong hyperfine interaction with nearby nuclear spins and ignore electric noises, we consider molecular clock sensors with transverse ZFSs that are not sensitive to magnetic noises and signals in linear order at clock transitions. The sensitivity enhancement is caused by two factors, strong continual RF driving turns the magnetic signal dependence from second-order to linear order and MW DD control leads to  $T_{2}$ exceeds $T_{2}^{*}$ \cite{sensitivity,Rondin2014magnetometry}. Therefore, the sensitivity of clock sensors is enhanced by $g_{ac}\sqrt{T_{2}^{*}}/E_{x}\sqrt{T_{2}}$.
A general and robust DD control method is proposed to further extend the spin coherence of these sensors, such as strained NV centers, defects in SiC, and optically addressable molecular spins. Our method focuses on electric noises and/or second-order magnetic noise. More importantly, the transverse ZFS acts as a frequency mixer and is controllable through symmetry, local strain, and the electric field, enabling a broadened sensing bandwidth. Although we consider zero magnetic field here, our method is not limited to absolute zero field but to the condition $\gamma_e B \ll E_x$.

In summary, we present a novel control method for zero-field magnetometry using a molecular clock sensor in the presence of a transverse ZFS. The main idea is to combine RF driving and DD pulse control, with continual RF driving resonant with the transverse ZFS, making the sensors sensitive to weak AC magnetic fields. This combination is robust to environmental noise and control errors, resulting in a long decoherence time and enabling the investigation of new systems and dynamics previously inaccessible at zero field.

\section{Acknowledgements}
J.W. Jiang and Q. Chen are supported by Natural Science Foundation of China (Grants No. 12247105,12375012), Hunan provincial major sci-tech program  (Grants No. 2023zk1010,  XJ2302001).

%M.B. Plenio was supported by the European Research Council Synergy Grant HyperQ (grant no. 856432) and the BMBF Zukunftscluster QSense: Quantensenoren f{\"u}r die biomedizinische Diagnostik (QMED) (grant no 03ZU1110FF).

\appendix
\section{The Hamiltonian of a molecular clock sensor}
We consider a molecular clock sensor, with spin $S = 1$ in the presence of a transverse zero field splitting (ZFS), which is described by the Hamiltonian
\begin{eqnarray}
H&=&D S_{z}^{2}+\delta_{B_{z}}S_{z}\\
&+&E_{x}(S_{x}^{2}-S_{y}^{2})+E_{y}(S_{x}S_{y}+S_{y}S_{x}),\notag
\end{eqnarray}
in which, $D$ is the longitudinal ZFS, $\delta_{B_{z}}$ is induced by Zeeman splitting due to the external magnetic field and couplings to the surrounding nuclear spins, the $E_{x}$ ($E_{y}$) is the $x$ ($y$) component of the transverse ZFS. According to the Eq. (1) , the eigenenergies and eigenstates of the molecular clock sensor are given by
\begin{eqnarray}
E_{+}^{\delta}&=&D+\sqrt{E^{2}_{x}+E^{2}_{y}+(\delta_{B_{z}})^{2}}, \\ \notag
E_{-}^{\delta}&=&D-\sqrt{E^{2}_{x}+E^{2}_{y}+(\delta_{B_{z}})^{2}}, \\ \notag
E_{0}^{\delta}&=&0,\notag
\end{eqnarray}
and
\begin{eqnarray}
|\psi_{+}\rangle&=&\cos\frac{\theta}{2}|+1\rangle+e^{i\phi}\sin\frac{\theta}{2}|-1\rangle,\\
|\psi_{-}\rangle&=&\sin\frac{\theta}{2}|+1\rangle-e^{i\phi}\cos\frac{\theta}{2}|-1\rangle,\notag\\
|\psi_{0}\rangle&=&|0\rangle,\notag
\end{eqnarray}
in which, the relevant angles satisfy the condition $\cos\theta=\frac{\delta_{B_{z}}}{\sqrt{E^{2}_{x}+E^{2}_{y}+(\delta_{B_{z}})^{2}}}$ and $\cos\phi=\frac{E_{x}}{\sqrt{E^{2}_{x}+E^{2}_{y}}}$.

We assume the transverse ZFS $E_{\perp}=\sqrt{E^{2}_{x}+E^{2}_{y}}=E_{x}$ by defining the $x$ axis to pass through the molecular clock sensor's transverse direction, the Hamiltonian is then represented by
\begin{eqnarray}
H^{'}&\approx&DS_{z}^{2}+\delta_{B_{z}}S_{z}
+E_{x}(S_{x}^{2}-S_{y}^{2}).
\end{eqnarray}
By considering a transverse ZFS far larger than the Zeeman splitting and the hyperfine coupling between the molecular clock sensor and nuclear spins, i.e., $\delta_{B_{z}}\ll E_x$, the Hamiltonian is simplified as
\begin{eqnarray}\label{A5}
H^{''}&\approx&DS_{z}^{2}+E_{x}(S_{x}^{2}-S_{y}^{2}),
\end{eqnarray}
where the main energy splitting is dominated by transverse ZFS at clock transition, the eigenenergy and eigenstates become
\begin{eqnarray}
E_{+}&=&D+E_{x},
\qquad|+\rangle=\frac{1}{\sqrt{2}}(|+1\rangle+|-1\rangle),\\
E_{-}&=&D-E_{x},
\qquad|-\rangle=\frac{1}{\sqrt{2}}(|+1\rangle-|-1\rangle),\notag\\
E_{0}&=&0,\qquad|0\rangle.\notag
\end{eqnarray}
%

%%%%%%%%%%%%%%%%%%%%%%%%%%%%%
\begin{figure}[ht]
\center\includegraphics[scale=0.52]{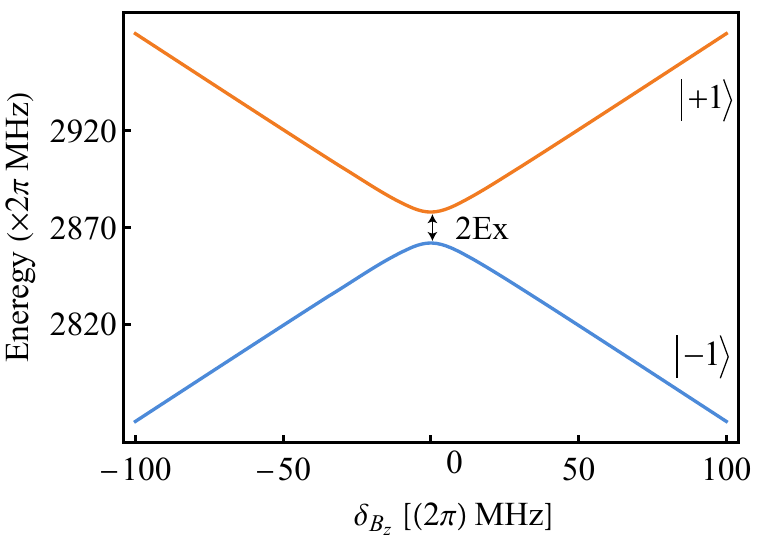}
\caption{Energy level diagram of NV center varying with magnetic field $\delta_{B_{z}}$, the states $|\pm\rangle$ are separated by transverse ZFS under zero-field, i.e., $\delta_{B_{z}}=0$.}
\label{energy}
\end{figure}
%%%%%%%%%%%%%%%%%%%%%%%%%%%%%

The recent reported transverse ZFS can be several MHz for shallow NV centers in diamond or more than 10 MHz in nanodiamonds \cite{nanodiamonds} due to local strains near the surface, could reach tens of MHz or be larger than 100 MHz for NV centers in SiC \cite{luo2023fabrication}, and GHz for molecular spins \cite{bayliss2022enhancing}. Take an nitrogen vacancy (NV) center as an example, suppose transverse ZFS $E_{x}=(2\pi) 8$ MHz and energy levels vary with the external magnetic in $|\pm1\rangle$ states is shown in Fig. \ref{energy}, one can find that the energy splitting is dominated by $2E_{x}$ between the states $|\pm\rangle$ at clock transition when $\delta_{B_{z}}\ll E_{x}$.

\section{The effective Hamiltonian engineering}
Continual radio-frequency (RF) driving and microwave (MW) pulse control are applied to the system, where the total Hamiltonian is described by
\begin{eqnarray}\label{HT}
H_{T}&=&DS_{z}^{2}+(E_{x}+\delta_{E})(S_{x}^{2}-S_{y}^{2})+\omega_{s}(t)S_{z}\\
&+&\delta_{B_{z}}S_{z}+\delta_{B_{x}}S_{x}+\delta_{B_{y}}S_{y}\notag\\
&+&\Omega_{rf}(1+\eta_{r})\cos(\omega_{rf}t)S_{z}\notag\\
&+&\Omega(t)(1+\eta_{m})\cos(\omega t+\varphi)S_{x},\notag
\end{eqnarray}
in which, $\omega_{s}(t)=g_{ac}\cos(\omega_{ac}t)$ is AC signal, $\Omega_{rf}$ and $\omega_{rf}$ are Rabi and driving frequency of RF field, $\Omega(t)$ and $\omega$ are time dependent Rabi frequency and driving frequency of MW pulse, $\varphi$ is the phase of MW driving. The system is subject to magnetic fluctuations ($\delta_{B_{x}}$, $\delta_{B_{y}}$, $\delta_{B_{z}}$), electric fluctuations ($\delta_{E}$), RF driving fields relative error ($\eta_{r}$) and MW pulse control relative error ($\eta_{m}$).

Firstly, we consider a sufficiently strong RF field is continually driven along the sensor's orientation $z$-direction in Eq. (\ref{HT}), the Hamiltonian is given by
\begin{eqnarray}\label{NV}
H_{sc}&=&D S_{z}^{2}+(E_{x}+\delta_{E})(S_{x}^{2}-S_{y}^{2})\\
&+&\delta_{B_{x}}S_{x}+\delta_{B_{y}}S_{y}+\delta_{B_{z}}S_{z}\notag\\
&+&\Omega_{rf}(1+\eta_{r})\cos(\omega_{rf}t)S_{z}.\notag
\end{eqnarray}
Go to rotating frame with respect of $H_{0}=DS_{z}^{2}+E_{x}(S_{x}^{2}-S_{y}^{2})$, the RF control is to be on resonance of the energy between eigenstates $|\pm\rangle$ with $\omega_{rf}=2E_{x}$, we have the Hamiltonian working in $\{|+\rangle,|0\rangle,|-\rangle\}$ space,
\begin{widetext}
\begin{eqnarray}\label{eff}
H_{sc}^{'}&=& \left(
            \begin{array}{ccc}
         \delta_{E} &  \delta_{B_{x}}e^{i(D+E_{x})t} & \Omega_{rf}(1+\eta_{r})/2 + \delta_{B_{z}}\cos(2E_{x}t)\\
              \delta_{B_{x}}e^{-i(D+E_{x})t}  & 0 & i\delta_{B_{y}}e^{-i(D-E_{x})t}  \\
               \Omega_{rf}(1+\eta_{r})/2+ \delta_{B_{z}}\cos(2E_{x}t) & -i\delta_{B_{y}}e^{i(D-E_{x})t}  & -\delta_{E} \\
            \end{array}
          \right).
\end{eqnarray}
\end{widetext}
It is shown that the system is first-order insensitive to magnetic field noises since $D+E_{x}\gg\delta_{B_{x}}$, $D-E_{x}\gg\delta_{B_{y}}$ and $2E_{x}\gg\delta_{B_{z}}$. All the magnetic fluctuations are suppressed in the first order at clock transitions. One can calculate the second-order magnetic fluctuations proportional to $\frac{\delta^{2}_{B_{x}}}{2(D+E_{x})}$, $\frac{\delta^{2}_{B_{y}}}{2(D-E_{x})}$ and $\frac{\delta^{2}_{B_{z}}}{4E_{x}}$. Because these second-order fluctuations have similar effects as $\delta_{E}$, it is reasonable to eliminate magnetic fluctuations and include $\delta_{E}$ instead both for electrically neutral molecular spins and color centers in diamond or SiC where $\delta_{E}$ is mainly determined by electric fluctuations. Therefore the total Hamiltonian from Eq. (\ref{HT}) becomes
\begin{eqnarray}\label{H}
H_{T}^{'}&=&DS_{z}^{2}+(E_{x}+\delta_{E})(S_{x}^{2}-S_{y}^{2})+\omega_{s}(t)S_{z}
\\&+&\Omega_{rf}(1+\eta_{r})\cos(\omega_{rf}t)S_{z}\notag\\
&+&\Omega(t)(1+\eta_{m})\cos(\omega t+\varphi)S_{x}.\notag
\end{eqnarray}
We move to the interaction picture, the rotating frame is defined as $H_{0}=DS_{z}^{2}+E_{x}(S_{x}^{2}-S_{y}^{2})$, then the effective Hamiltonian is
\begin{widetext}
\begin{gather}\label{int}
H_{int}=U^{\dag}_{0}(t)HU_{0}(t)-iU_{0}(t)\frac{\partial U^{\dag}_{0}(t)}{\partial t},\\
=\left(
     \begin{array}{ccc}
       [\Omega^{'}_{rf}\cos(\omega_{rf}t)+\omega_{s}(t)]\cos(2E_{x}t) & \frac{\sqrt{2}\Omega^{'}(t)}{2}\cos(\omega t+\varphi)e^{-i(D+E_{x})t} & i[\Omega^{'}_{rf}\cos(\omega_{rf}t)+\omega_{s}(t)]\sin(2E_{x}t)+\delta_{E} \\
        \frac{\sqrt{2}\Omega^{'}(t)}{2}\cos(\omega t+\varphi)e^{i(D+E_{x})t} & 0 &  \frac{\sqrt{2}\Omega^{'}(t)}{2}\cos(\omega t+\varphi)e^{i(D+E_{x})t} \\
        -i[\Omega^{'}_{rf}\cos(\omega_{rf}t)+\omega_{s}(t)]\sin(2E_{x}t)+\delta_{E} &  \frac{\sqrt{2}\Omega^{'}(t)}{2}\cos(\omega t+\varphi)e^{-i(D+E_{x})t} & -[\Omega^{'}_{rf}\cos(\omega_{rf}t)+\omega_{s}(t)]\cos(2E_{x}t) \\
     \end{array}
   \right),\notag
\end{gather}
\end{widetext}
where, the evolution propagator $U_{0}(t)=\exp(-iH_{0}t)$, the $\Omega^{'}_{rf}=\Omega_{rf}(1+\eta_{r})$ and $\Omega^{'}(t)=\Omega(t)(1+\eta_{m})$.

We assume $|\omega_{ac}-2E_{x}|\ll E_{x}\ll|\omega_{ac}+2E_{x}|$, $\omega_{ac}$, $E_{x}>0$ and the RF resonance condition $\omega_{rf}=2E_{x}$, $\frac{1}{2}\Omega_{rf}\ll2E_{x}$,  and rotating-wave approximation (RWA) is applied. We assume that the MW driving frequency is on resonance with $\omega=D+E_{x}$ and $\varphi=0$, and the effective Hamiltonian is given by
\begin{widetext}
\begin{eqnarray}\label{inteff}
H_{eff}&\approx& \left(
            \begin{array}{ccc}
              \frac{1}{2}[\Omega_{rf}(1+\eta_{r})+g_{ac}\cos(\Delta\omega t)] &  \frac{\sqrt{2}}{4}(1+\eta_{m})\Omega(t) & \delta_{E} \\
               \frac{\sqrt{2}}{4}(1+\eta_{m})\Omega(t) & 0 &  \frac{\sqrt{2}}{4}(1+\eta_{m})\Omega(t) \\
              \delta_{E} &  \frac{\sqrt{2}}{4}(1+\eta_{m})\Omega(t) & -\frac{1}{2}[\Omega_{rf}(1+\eta_{r})+g_{ac}\cos(\Delta\omega t)] \\
            \end{array}
          \right),
\end{eqnarray}
\end{widetext}
where the detuning $\Delta\omega=\omega_{ac}-2E_{x}$. All of our numerical simulations in section \uppercase\expandafter{\romannumeral3} and \uppercase\expandafter{\romannumeral4} are based on the above Hamiltonian in Eq. (\ref{inteff}). The free evolution without MW pulse control of the system is described by following Hamiltonian
\begin{eqnarray}
H_{free}^{'}&\approx&\frac{1}{2}[\Omega_{rf}(1+\eta_{r})+g_{ac}\cos(\Delta\omega t)]S_{z}\notag\\
&+&\delta_{E}(S_{x}^{2}-S_{y}^{2}),
\end{eqnarray}
thus, considering in $|\pm1\rangle$ subspace, the free evolution Hamiltonian can be rewritten as
\begin{eqnarray}\label{free}
H_{free}\approx\Omega_{rf}(1+\eta_{r})\sigma_{z}+g_{ac}\cos(\Delta\omega t)\sigma_{z}+2\delta_E\sigma_{x},
\end{eqnarray}
in which Pauli operators are defined as $\sigma_{z}=\frac{1}{2}(|+1\rangle\langle+1|-|-1\rangle\langle-1|)$ and $\sigma_{x}=\frac{1}{2}(|+1\rangle\langle-1|+|-1\rangle\langle+1|)$, the RF driving is not only suppress transverse noise but also make the system sensitive to weak magnetic field signal.

MW fields are applied to control the molecular clock sensor in \{$|0\rangle,|+\rangle$\} basis under zero-field while the system is applied continuouslly RF field driving, according Eq. (\ref{inteff}) the effective Hamiltonian is given by
\begin{eqnarray}\label{d1}
H_{d}&\approx&\frac{\Omega(1+\eta_{m})}{2}(|0\rangle\langle
+|+|+\rangle\langle0|)\\
&+&\frac{\Omega_{rf}(1+\eta_{r})}{2}(|-\rangle\langle+|+|+\rangle\langle -|)\notag\\
&+&\delta_E(|+\rangle\langle +|-|-\rangle\langle -|),\notag
\end{eqnarray}
since $D\gg E_{x}$, the high frequency term $(D-E_{x}-\delta_{E})|-\rangle\langle-|$ is neglected during the calculations. According to Eq. (\ref{d1}), the evolution propagator of microwave pulse is given by
\begin{eqnarray}\label{pulse2}
U_{p}=\exp\left[-i\int_{0}^{T_{p}}H_{d}dt\right].
\end{eqnarray}
If transverse noises $\delta_{E}$ and RF control errors are negligible ($\delta_{E},\eta_{m},\eta_{r}\ll\Omega_{rf}\ll\Omega$), the strength of pulse is defined as $T_{p}\approx\frac{2\pi}{\Omega}$ (note that $\int_{0}^{T_{p}}\Omega/2=2\pi$), after such a $2\pi$ pulse, the state $|+\rangle$ evolves as $\exp\left[-i\int_{0}^{T_{p}}H_{d}dt\right]|+\rangle$=$-|+\rangle$ when the state $|-\rangle$ is unchanged. The flipping between $|\pm1\rangle$ and $-|\pm1\rangle$ is realized. Thus our $2\pi$ pulse working in \{$|0\rangle,|+\rangle$\} subspace plays the same role as a $\pi$ pulse working in \{$|+1\rangle,|-1\rangle$\} subspace.

Now one can find that how our scheme works. Our sensor is working in  \{$|+1\rangle,|-1\rangle$\} subspace, the system is initialized in state $|+\rangle=\frac{1}{\sqrt{2}}(|+1\rangle+|-1\rangle)$, the AC magnetic signal are accumulated during free evolution when the continual RF driving is applied. The MW pulses are used to have a $\pi$ pulse working in \{$|+1\rangle,|-1\rangle$\} subspace through a $2\pi$ pulse working in \{$|0\rangle,|+\rangle$\} subspace. The MW dynamic decoupling (DD) pulses provides an access to decouple the system from environmental noises. Then, we calculate numerically the propagator
\begin{eqnarray}\label{propagator}
U^{(n)}=U_{\tau}\underbrace{U_{p}U_{2\tau}\ldots U_{2\tau}U_{p}}_{n}U_{\tau},
\end{eqnarray}
where $U_{\tau}$ ($U_{2\tau}$) is free evolution with only continuous RF field control and $U_{p}$ the pulse controls.

\section{The dynamical decoupling mechanism}
%%%%%%%%%%%%%%%%%%%%%%%%%%%%%
\begin{figure}[ht]
\center\includegraphics[scale=0.45]{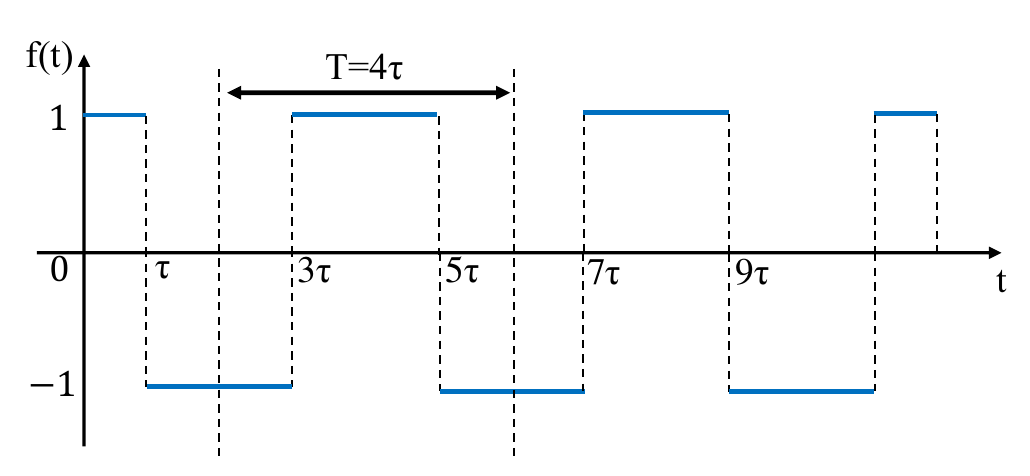}
\caption{The modulation function of rectangular pulse sequence flips between the $+1$ and $-1$. }
\label{filter}
\end{figure}
%%%%%%%%%%%%%%%%%%%%%%%%%%%%%
To understand the resonance condition between the pulses period and target frequency, the action of any pulse sequences can be described by the filter function $F(\omega)$ and the response function $f(t)$, which represent changes in the frequency and time domain separately. The response function $f(t)$ altenatively has a value of $+1$ or $-1$ between every two $2\pi$ pulses, see Fig. \ref{filter}. For example, the response function of rectangular pulse sequence is given by
\begin{eqnarray}
f(t)&=&\begin{cases}
1,&\text{$1<t<\tau$}\\
-1,&\text{$\tau<t<3\tau$}
\end{cases}
\end{eqnarray}
thus, the effective Hamiltonian of whole system with DD control is as follows
\begin{eqnarray}\label{total}
H(t)\approx f(t)[\Omega_{rf}(1+\eta_{r})\sigma_{z}+g_{ac}\cos(\Delta\omega t)\sigma_{z}+2\delta_{E}\sigma_{x}].\notag\\
\end{eqnarray}
Here, the associated filter function and Fourier coefficient is
\begin{eqnarray*}
F(t)&=&\sum_{n=1}^{\infty}a_{n}\cos\left(\frac{2n\pi t}{T}\right),\\
a_{n}&=&\frac{2}{T}\int_{0}^{T}f(t)\cos\left(\frac{2n\pi t}{T}\right)dt.
\end{eqnarray*}
Through calculation
\begin{eqnarray*}
a_{n}&=&\frac{4}{n\pi}\left[(-1)^{n+1}+1\right],n=1,2,3.....\\
F(t)&=&\sum_{n=1}^{\infty}\frac{4}{n\pi}\left[(-1)^{n+1}+1\right]\cos\left(\frac{2n\pi t}{T}\right),n=1,2,3...
\end{eqnarray*}
The frequency is defined as $\omega_{T}=2\pi/T$, where $T$ is the period of filter function, then
\begin{eqnarray*}
F(\omega_{T},t)=\sum_{n=1}^{\infty}\frac{4}{n\pi}\left[(-1)^{n+1}+1\right]\cos\left(n\omega_{T}t\right),n=1,2,3...
\end{eqnarray*}
in the case $n=1$,
\begin{eqnarray*}
F(\omega_{T},t)=\frac{8}{\pi}\cos\left(\omega_{T}t\right).
\end{eqnarray*}
Thus, the total Hamiltonian in Eq. (\ref{total}) could be rewritten as
\begin{eqnarray}
H^{'}(t)&=&\frac{8}{\pi}\cos(\omega_{T}t)[\Omega_{rf}(1+\eta_{r})\sigma_{z}\\
&+&g_{ac}\cos(\Delta\omega t)\sigma_{z}+2\delta_{E}\sigma_{x}],\notag
\end{eqnarray}
if the target frequency is resonant with the modulation frequency, i.e., $\omega_{T}=\Delta\omega=\frac{2\pi}{T}$, the target AC signal is detected with an accumulated phase $\eta(t)\equiv\int_{0}^{t}g_{ac}|\cos(\Delta\omega t)|$ under this periodic filter function. The time interval $\tau$ of pulse in DD control is related to the period of filter function, where $\tau=\frac{T}{4}=\frac{\pi}{2\Delta\omega}$.

\section{Robustness}
In the numerical simulation these noise ($\delta_{E}$, $\eta_{m}$, $\eta_{r}$) are modeled as an Ornstein-Uhlenbeck (OU) process \cite{wang1945theory,gillespie1996mathematics,gillespie1996exact} with a zero expectation value $\langle\delta(t)\rangle=0$. For example, we give the dynamical noise of electronic $\delta_{E}=\delta_{E}(t)$ modeled as an OU process with a zero expectation value $\langle\delta_{E}(t)\rangle=0$, the exact algorithm of OU process is
\begin{eqnarray}
\delta_{E}(t+\Delta t)=\delta_{E}(t)e^{-\Delta t/\tau_{c}}+\widetilde{n}\sqrt{\frac{c\tau}{2}(1-e^{-2\Delta t/\tau})},
\end{eqnarray}
where, $\widetilde{n}$ is a unit Gaussian random number. When the noise $\delta_{E}(t)$ satisfies diffusion constant $c=4/(T_{2}^{*}\tau)$ and the correlation time is set to $\tau_{c}=20$ $\mu$s, the crucial parameter is $T_{2}^{*}$ which determines the decohenrence time of clock sensor center \cite{trusheim2014scalable}. We average results that are calculated by performing the simulation 1200 times for different noise realizations. The MW pulse fluctuations are modeled by uncorrelated OU processes with the same correlation time $\tau_{m}=500$ $\mu$s and a relative amplitude error $\eta_{m}=0.5\%$ with the corresponding diffusion constant $c_{m}=2\eta_{m}^{2}\Omega^{2}/\tau_{m}$ \cite{genov2019mixed}. The RF driving fluctuations are defined in a similar way as for MW pulse.

%%%%%%%%%%%%%%%%%%%%%%%%%%%%%
\begin{figure*}[ht]
\center\includegraphics[scale=0.52]{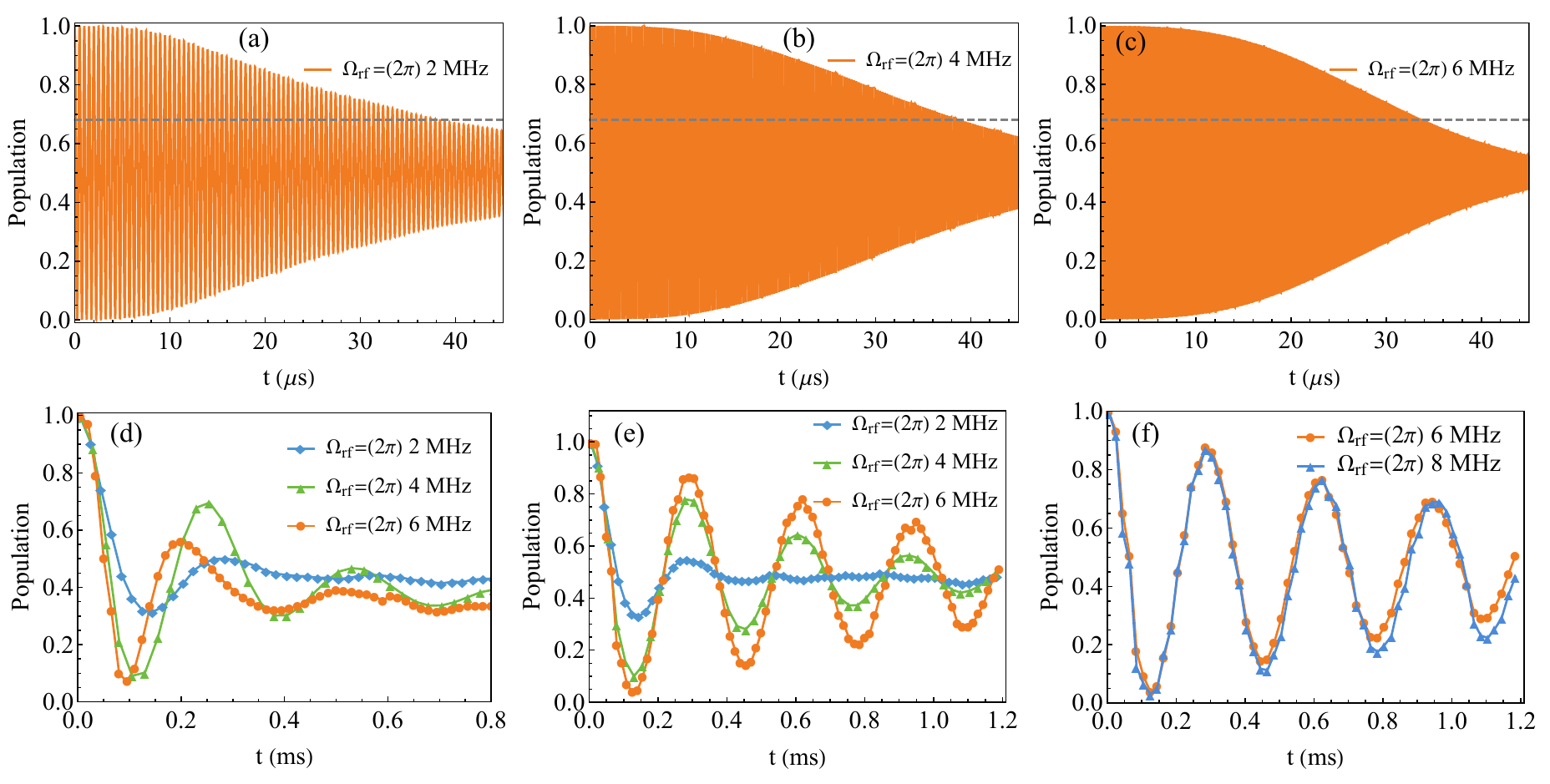}
\caption{Simulations under different pulse sequences and RF controls. All results use MW pulse control with frequency $\Omega=(2\pi)$ $40$ MHz and the resonance condition $\Delta\omega=(2\pi)$ $0.1$ MHz. The coupling of target AC field is $g_{ac}=(2\pi)$ $5$ kHz, the expectation coherence time $T_{2}^{*}=1.8$ $\mu$s, the fluctuations of RF and MW amplitude both are $0.5\%$. Evolutions of $|+\rangle$ state are shown for (a) free evolution when the Rabi frequency of RF $\Omega_{rf}=(2\pi)$ $2$ MHz, (b) free evolution with RF frequency $\Omega_{rf}=(2\pi)$ $4$ MHz, (c) free evolution with $\Omega_{rf}=(2\pi)$ $6$ MHz, (d) CPMG sequences with $\Omega_{rf}=(2\pi)$ $(2,4,6)$ MHz, (e) LDD8b sequences with $\Omega_{rf}=(2\pi)$ $(2,4,6)$ MHz, (f) LDD8b sequences with $\Omega_{rf}=(2\pi)$ $(6,8)$ MHz respectively.}
\label{LDD}
\end{figure*}
%%%%%%%%%%%%%%%%%%%%%%%%%%%%%
%can resist the control errors up to $2\pi\times6$ MHz, but there is no more benefit when $\Omega_{rf}=2\pi\times8$ MHz.

In order to illustrate the robustness of our scheme, we give detail calculations of errors for the free evolution. We have given the effective free evolution Hamiltonian $H_{free}$ and the evolution propagator $U_{\tau}$ is defined as $U_{\tau}=\exp[-iH_{free}\tau]$. If the Rabi frequency of RF field satisfies $\Omega_{rf}=2\pi\times2n$ ($n=1,2,3...$), $\tau=2.5$ $\mu$s and $\eta_{r}$ is small enough to be ignored, the $U_{\tau}$ is approximated as
\begin{eqnarray}\label{error}
U_{\tau}&\approx& \textit{I}-i\tau\frac{\delta^{2}_{E}}{\Omega_{rf}}\sigma_{z}-i\tau\frac{2\delta^{3}_{E}}{\Omega^{2}_{rf}}\sigma_{x},
\end{eqnarray}
where, $\textit{I}$ is the unit operator representing evolution without errors, $\frac{\delta^{2}_{E}}{\Omega_{rf}}$ is the second order error induced by $\delta_{E}$, $2\tau$ is pulse interval with $\tau=\frac{T}{4}$. It should be noted that we haven't considered the contribution of the target AC field signal $g_{ac}\cos(\Delta\omega t)$ because it is weak and we focus on the error accumulation here.

Assuming coherence evolution during a sequence of pulses with noise detuning, the propagator is calculated by Eq. (\ref{propagator}),
for example, we consider only two pulses, i.e., $n=2$, the evolution propagator is given by assuming the MW $\pi$ pulses are perfectly applied
\begin{eqnarray}
U^{(2)}=\left(
          \begin{array}{cc}
            -1+\frac{32\delta_{E}^{6}\tau^{2}}{\Omega_{rf}^{4}} & i\frac{8\delta_{E}^{3}\tau}{\Omega_{rf}^{2}} \\
            i\frac{8\delta_{E}^{3}\tau}{\Omega_{rf}^{2}} & -1+\frac{32\delta_{E}^{6}\tau^{2}}{\Omega_{rf}^{4}} \\
          \end{array}
        \right).
\end{eqnarray}
We have found in our calculations that it is useful to eliminate the second order errors $\tau\frac{\delta^{2}_{E}}{\Omega_{rf}}$ induced by electric noise $\delta_{E}$ via a second $2\pi$ pulse and the decoherence time $T_{2}$ of NV center depends on the third order as $\frac{2\delta^{3}_{E}}{\Omega^{2}_{rf}}$.

\end{document}